\documentclass[twocolumn,showpacs,preprintnumbers,amsmath,amssymb,showkeys]{revtex4-1}

\usepackage{graphicx}
\usepackage{dcolumn}
\usepackage{bm}
\usepackage{bezier,color}

\newcommand{\bea}{\begin{eqnarray}}
\newcommand{\eea}{\end{eqnarray}}
\newcommand{\nn}{\nonumber}

\newcommand{\la}{\langle}
\newcommand{\ra}{\rangle}

\newcommand{\f}{\frac}

\begin{document}

\title{Sources of stochasticity in constitutive and autoregulated gene expression}
\author{Rahul Marathe$^{(1)}$, David Gomez$^{(1,2)}$ and Stefan Klumpp$^{(1)}$}
\affiliation{$^{(1)}$Max Planck Institute of Colloids and Interfaces, Science Park Golm, 14424 Potsdam,
Germany \\ 
$^{(2)}
$Department of Physics, Freie Universit{\"a}t Berlin, Arnimallee 14,  14195 Berlin, Germany
\email{rahul.marathe@mpikg.mpg.de}}
\date{\today}

\begin{abstract}
Gene expression is inherently noisy as many steps in the read-out of the genetic information are stochastic. To disentangle the effect of different sources of stochasticity in such systems, we consider various models that describe some processes as stochastic and others as deterministic. We review earlier results for unregulated (constitutive) gene expression and present new results for a gene controlled by negative autoregulation with cell growth modeled by linear volume growth. 
\end{abstract}
\maketitle

\section{Introduction}
\label{sec:Introduction}
Mathematical and physical methods and concepts are increasingly used in the life sciences.
For example,  the dynamics of gene regulatory circuits 
is often studied by describing these circuits with simple but non-trivial mathematical models \cite{Hasty,Bintu2005,Guido06,Wall}. An important issue is then the choice of an appropriate level of mathematical description. Cells are very dynamic and often adapt to the external
conditions by changing their global properties, which may  in turn affect the function of genetic circuits hosted by that cell \cite{Klumpp09,TanYou,Scott,KlumppPlasmid}. 
Thus one needs to address the question of how one can mathematically model such a dynamic system, at least under 
constant external conditions. 
Moreover, every individual cell grows and divides while the circuits it contains perform their programmed functions. 
Many models use a mean-field-like description averaging over these processes, but it is often not clear how accurate such approximations are. For example, the gene copy number is often described by an average and the actual
doubling of the gene during the cell division cycle is not considered. 

Yet another issue is  whether the description of such a regulatory circuit should be deterministic or stochastic. Many important molecules are present in a cell in low copy numbers. Hence, fluctuations can be expected to be important,  so that a stochastic description of gene expression is necessary \cite{BergJTB78,AdamsPNAS97,ElowitzSci02}. 
These effects, which have been studied theoretically for a long time \cite{BergJTB78,KoJTB91,YcartTPB95,CookPNAS95,
HastyPNAS2000,OudenaardenPNAS01,SwainPNAS2002,PaulssonReview,Scott07}, have recently become accessible to direct quantitative experiments thanks to the development of single cell approaches  \cite{ElowitzSci02,Golding,YuSci06,xie}.

During the division cycle of a cell, stochasticity
arises from different sources and at different points, namely from the inherent stochasticity of the synthesis of proteins (which occurs throughout the division cycle) and from the partitioning of the protein molecules among the daughter cells during cell division (an approximately instantaneous event).  An obvious question that arises in this context is whether there is a dominant source
of noise? This question is related to the problem of which mathematical description is most appropriate: Which sources of noise need to be taken into account for a minimal, but realistic description? Do different descriptions of the noise, with or without explicit volume growth, with explicit or implicit 
cell division etc. lead to approximately the same predictions or are there considerable differences between these descriptions concerning the  noise that is generated? In a recent study \cite{MaratheJSP11}, we have addressed some of these issues by considering various simple models that include or exclude certain sources of noise. The comparison of these results has shown that often there is no dominant source of noise, i.e.\ that different sources contribute comparably (an exception is so-called bursty protein synthesis: if many proteins are produced from relatively rare transcription events, this bursting is clearly the dominant source of noise). The absence of a dominant noise source means that on the one hand, all sources have to be included for accurate results, but, on the other hand, also that omitting any of those sources will still lead to fluctuations of the same order of magnitude. 

In our previous study, these questions have been studied for unregulated genes. Here we extend our approach to regulated genes. We focus on a simple, but important regulatory system, namely negative autoregulation, where the protein product of a gene controls the read-out of that gene, such that large concentrations of the protein suppress further synthesis of that protein \cite{SavageauNat74,Becskei2000,OudenaardenPNAS01}. Fluctuations arising from both sources we consider (stochastic synthesis and stochastic partitioning during cell division) are found to be suppressed substantially by the negative feedback. A complication that arises generally for regulated genes is that regulation depends on protein \emph{concentrations}, which in turn depend on the cell volume. This means that the growth of cellular volume needs to be taken into account explicitly, which was not necessary for unregulated genes that could be described fully by the \emph{number} of protein molecules per cell. 

%

The paper is organized as follows: 
In section \ref{summary}, we review some key results from our previous study \cite{MaratheJSP11} comparing different sources of noise for unregulated gene expression. An alternative analytical derivation of one central result is presented in the appendix. In section \ref{model},  this type of analysis is extended to a gene controlled by negative autoregulation. We end with some  concluding remarks.

\section{Sources of stochasticity for constitutive gene expression}
\label{summary}
Recently we studied different models for the stochastic gene expression of an unregulated (constitutively expressed gene) in order to disentangle different sources of (intrinsic) stochasticity \cite{MaratheJSP11}. 
Stochasticity arises from the process of protein synthesis  and also from degradation, if the proteins are unstable. When
a cell divides, the partitioning of proteins among daughter cells also generates fluctuations. In our recent work \cite{MaratheJSP11}, we analyzed these different sources in a systematic way to see which sources contribute to the observed noise and whether there is a dominant source. In this section, we briefly summarize some key results obtained with these models.  

Protein synthesis is a two-step process consisting of transcription and translation. In the first step, a gene sequence is transcribed 
into mRNA and then it is translated by ribosomes to produce proteins. If $M$ and $P$ represent the mRNA and protein copy numbers, respectively,  their time
evolution is described by:
\begin{eqnarray}
\dot{M} & = & \alpha_m g -\beta_m M \nn \\
\dot{P} & = & \alpha_pM-\beta_p P, \label{eq:P_M}
\end{eqnarray}
where $\alpha_m$, $\alpha_P$ and $\beta_m$, $\beta_p$ are synthesis and degradation rates of mRNAs and proteins respectively. $g$ is the
gene copy number.  In bacteria, proteins are often stable (with lifetimes long compared to the generation time $T$) \cite{NathJBC70}. Then the degradation rate in Eq.\ \ref{eq:P} is an effective degradation rate representing dilution by cell growth and division with $\beta=\ln2/T$. By contrast mRNA is typically rather short-lived with lifetimes in the range of a few minutes  \cite{Bernstein, xie} and one can approximate the equation for $M$ by its steady state,
$M=\alpha_m g/\beta_m$. In that case the above two-step process is reduced to an effective one-step process:
\bea
\dot{P} & = & \alpha g -\beta P, \label{eq:P}
\eea
with $\alpha=\alpha_p\alpha_m/\beta_m$. We would like to note that when mRNA is treated as a fast variable and considered to be in 
a steady state, one obtains a correct description of the average protein number, but the fluctuations are underestimated, in particular, if a single transcription event (or one mRNA molecule) gives rise to many copies of the protein. This effect, where the output of transcription is strongly amplified by translation, is known as bursty protein synthesis and will not be considered here. The reader is referred to ref. \cite{BergJTB78,OudenaardenPNAS01,MaratheJSP11} for discussions of this issue

\begin{figure}
\vspace{0.5cm}
\includegraphics[width=8cm, height=5cm,angle=0.0]{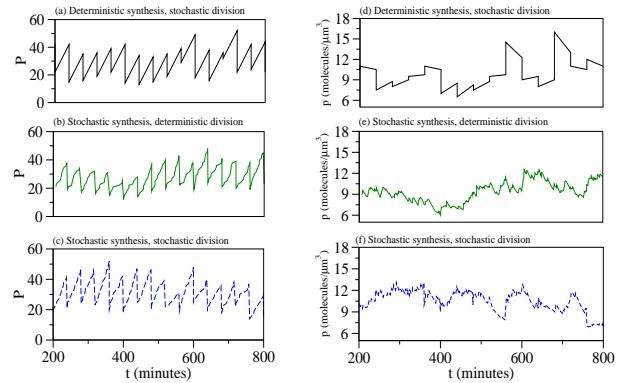}
\caption{Stochastic models of unregulated protein synthesis: (a)-(c) Trajectories of the protein copy number from stochastic simulations with 
stochastic synthesis, partitioning during cell division, or both, all with cell division modeled explicitly. (d)-(f) Corresponding concentrations of  the
protein calculated for a volume that increases linearly during the division cycle and is halved at multiples of the division time $T$. Here the cell volume 
does not affect the protein synthesis rate does. The parameter values used for these plots are $\alpha=0.5/$min, $T= 40$ min.}
\label{fig:stoch}
\end{figure}

\begin{figure}
\includegraphics[width=8cm, height=5cm,angle=0.0]{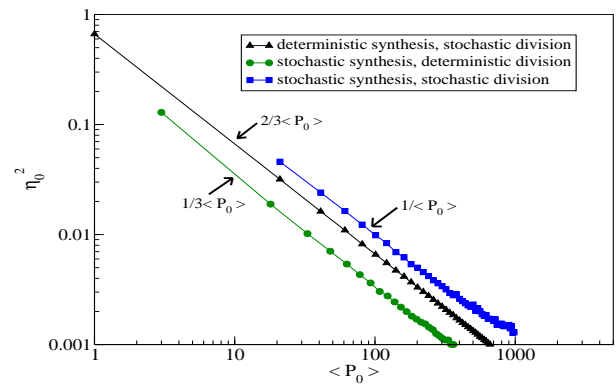}
\caption{Stochastic models of protein synthesis:  Noise strength $\eta^2$ as a function 
of the average protein copy number $\la P\ra$ (varied by varying the synthesis rate $\alpha$) for the different models (for the models 
with explicit cell division, averages over cell immediately after division are plotted, i.e. $\eta_0^2$ and $\la P_0\ra$). 
$T= 40$ min.}
\label{fig:etastoch}
\end{figure}

We now consider different models \cite{MaratheJSP11} that are  based on Eq.\ \ref{eq:P}, but describe cell division explicitly. In that case, the degradation rate for stable proteins is $\beta=0$ and the protein copy number per cell is divided by 2 at cell division. We start with a model where protein synthesis is described
deterministically, while proteins are distributed stochastically among the two daughter cells during cell division (we note that in all our models 
a cell divides in exactly two daughter cells and we look at only one lineage of cells; for some more complex cases see, e.g. \cite{AngeloJSP11}). So during division 
each protein molecule has a probability $r=1/2$ to go to either of the daughter cells.  
This means that in every generation a constant number $Q=\alpha T$ of protein molecules
are synthesized, but the initial protein number in each cycle fluctuates due to the stochastic division. Figure.\ \ref{fig:stoch}(a) shows a time series of 
such a process as obtained from simulations.  For this case we have obtained a number of analytical results \cite{MaratheJSP11}  using a method proposed in Ref. \cite{BrennerPRL07}. An alternative derivation based on generating functions is given in Appendix \ref{appA1}.  The average copy number after division and the variance of that number are found to be given by $ \la P_0\ra = Q =\alpha T$ 
and $\delta P_0^2=2Q/3$, respectively. Two commonly used characteristics of noise are the noise strength $\eta^2$ defined as
\begin{equation}
\label{def_eta}
\eta^2=\frac{\la (P-\la P\ra)^2\ra}{\la P \ra^2}
\end{equation}
and the Fano factor $F=\eta^2 \la P\ra$.  $\eta^2$ typically scales as $\eta^2\sim 1/\la P\ra$, so the latter parameter provides a 
characterization of the pre-factor of that scaling. For the case under consideration, we obtain 
\begin{equation}\label{eta:stochPart}
\eta_0^2=\f{2}{3\la P_0\ra}
\end{equation}
or $F_0=2/3$ (the index '0' in these expressions indicates that we have taken averages over a population of cells immediately after division).

In the complementary case, synthesis of proteins is stochastic and division deterministic. 
So when a cell divides each daughter cell gets exactly half of the available proteins as shown in figure \ref{fig:stoch}(b) (for odd number of protein $P$, 
we take the number after division to be either $(P+1)/2$ or $(P-1)/2$, each with probability $1/2$, thus leading to a minimal 
remnant of stochasticity in our otherwise deterministic description of cell division). We also assume the synthesis rate to be constant and do not explicitly describe  gene duplication. We then obtain 
\begin{equation}
\la P_0\ra=\alpha T,\qquad \delta P_0^2=\f{\alpha T}{3}\qquad{\rm{and}}\qquad \eta_0^2=\f{1}{3\la P_0\ra}.
\end{equation}
The last result implies that the Fano factor is $F_0=1/3$, which is just half of what we have seen for stochastic partitioning 
(Eq.\ \ref{eta:stochPart}).     

Finally, we combine both sources of stochasticity, thus synthesis as well as degradation of proteins occur stochastically (figure \ref{fig:stoch}c).  
Using again the method of Ref. \cite{BrennerPRL07}, we obtain 
\begin{equation}\label{combined}
\la P_0\ra=\alpha T,\qquad \delta P_0^2=\alpha T,\qquad {\rm and}\qquad \eta_0^2=1/\la P_0\ra.
\end{equation}
Points to be noted are: (i) Additive independent noise strengths ($\eta^2$). In our case, the noise in 
Eq.\ \ref{combined} is the sum of the noise components for stochastic partitioning ($2/\la 3 P_0\ra$) and from stochastic 
synthesis ($1/\la 3P_0\ra$). (ii) The contributions from both sources of noise are of the same  order of magnitude, implying that there is no dominant source of 
noise in this simple case. 

In figure \ref{fig:stoch}(d)-(f) we show time series for the concentrations of the protein for 
the three cases discussed above. The concentration fluctuates around its mean, and shows no systematic dependence on the cell division cycle. The latter observation arises form the fact that both the volume and the protein number increase (on average) linearly during the cycle. A systematic variation over the course of the division cycle is obtained if an explicit description of gene duplication is included or if the volume growth is not linear \cite{MaratheJSP11}.

\section{Protein synthesis with negative autoregulation}
\label{model}

\begin{figure}
\vspace{0.5cm}
\includegraphics[width=8cm, height=5cm,angle=0.0]{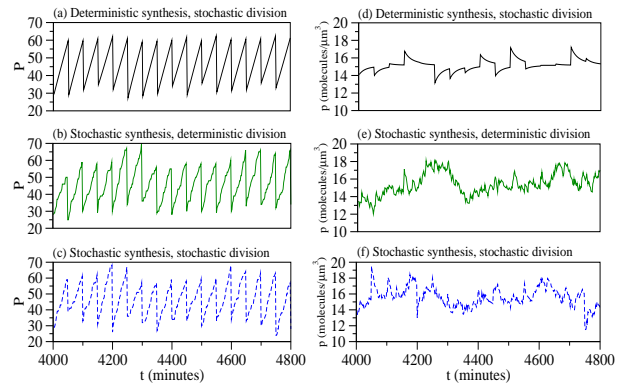}
\caption{Stochastic models of protein synthesis with negative auto-regulation: (a)-(c) Trajectories of the protein copy number from stochastic simulations with 
stochastic synthesis, cell division, or both, all with cell division modeled explicitly and linear volume growth. (d)-(f) Corresponding concentration of the protein. The parameter values used for these plots are 
$\alpha_0=2.0/min$, $V_0=2 \mu m^3$, $T= 50$ min, $K=10$ molecules$/\mu m^3$. }
\label{fig:autostoc}
\end{figure}
 
\begin{figure}
\hspace{-1.cm}
\includegraphics[width=8cm,height=6cm,angle=0.0]{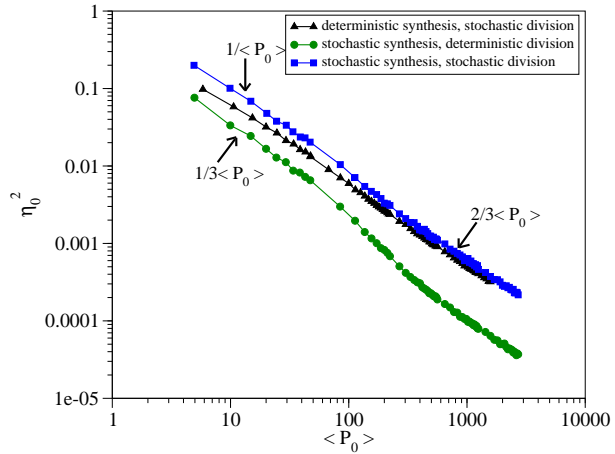}
\caption{ Noise strength $\eta_0^2$ as a function of the average protein copy number $\la P_0\ra$ 
(varied by varying the synthesis rate $\alpha_0$) for the different models with negative autoregulation. Averages are taken over cells immediately 
after cell division. The parameters values are $V_0=2 \mu m^3$, $T= 50$ min, $K=100$ molecules$/\mu m^3$.}
\label{fig:etaauto}
\end{figure}

Gene regulation is incorporated into models of the type of Eq.\ \ref{eq:P} via synthesis (or degradation) rates that depend on the concentration of a regulatory protein, for example a transcription factor. Here we consider one specific case, namely negative autoregulation, where the output of a gene (the protein product) modifies the read-out of that gene in such a way that  the synthesis  of a protein is suppressed by a high concentration of that protein \cite{ElowitzNat00, SimpsonPNAS03, Ozbudak02, OudenaardenPNAS01}. The dependence of the synthesis rate on the protein concentration $p=P/V$ is expressed by a so-called Hill function \bea
\alpha = \alpha(p)=\frac{\alpha_0}{1+(\frac{p}{K})^n }.
\label{autoP}
\eea
Here $\alpha_0$ is the maximal synthesis rate, $K$ is a concentration scale that defines which protein concentration is required to affect the synthesis rate (in the simplest case, it is given by the dissociation constant for binding of a transcription factor to its binding site on the DNA). $n$ is called the Hill coefficient, which describes the cooperativity of regulation and characterizes the steepness of the regulation function. In the following we take $n$ to be equal to 2.  

The synthesis rate $\alpha(p)$ in Eq.\ (\ref{autoP}) is time-dependent through both the protein copy number (which changes in discrete steps of synthesis and degradation) and the cell volume (which changes in a continuous fashion). So in contrast to the unregulated case discussed before, volume growth affects the dynamics of the protein synthesis process in the presence of (concentration-dependent) gene regulation.
In our case, volume growth is taken to be linear in time, starting from an initial volume $V_0$ directly after cell division and reaching $2V_0$ just before the next division. The growth is implemented via a discrete time step $\Delta t$ in which the volume increases by $\Delta V$. The volume is halved exactly at the division. We are again interested in the different sources
of stochasticity and consider the synthesis of protein and the partitioning of molecular
content to be either stochastic or deterministic.

We begin with the case where both synthesis and division are stochastic. 
The variation of the protein number versus time for this case is depicted in figure\ \ref{fig:autostoc}(c), the corresponding concentration
is shown in  figure \ref{fig:autostoc}(f). As before, we consider the dependence of the noise parameters $\eta$ on the average protein number. 
In this case $\eta_0^2$ follows a  $1/\la P_0\ra$-behavior for small $\la P_0\ra$, 
but crosses over to $2/3\la P_0\ra$ for large $\la P_0\ra$ (see figure.\ \ref{fig:etaauto} blue line with filled squares). 
The crossover occurs for values of $\la P_0\ra$ of the order of $KV_0$, i.e. it occurs when the autoregulation mechanism becomes 
important. For smaller $\la P_0\ra$ (or $\alpha_0$), the system behaves like an unregulated system with considerable fluctuations due to  
protein synthesis as well as division. For large $\alpha_0$ (or $\la P_0\ra$) autoregulation becomes active and suppresses protein number 
fluctuations, so that protein synthesis becomes approximately deterministic, but partitioning during division remains stochastic,  hence leading 
to the observed $2/3\la P_0\ra$-behavior.

Now we separate the two sources as we did before for the unregulated gene. In the first case, proteins are added deterministically and partitioned stochastically. 
Deterministic addition means integrating Eq.\ (\ref{autoP}) over a cycle. This number depends on the initial
protein number. The integration usually leads to a non integer value of the protein number, in such cases the remaining non integer part is interpreted probabilistically and it is added with a 
probability equal to the fractional part. A trajectory of the number of  molecules per cell for this case 
is depicted in figure \ref{fig:autostoc}(a) and the corresponding concentration 
is shown in figure \ref{fig:autostoc}(d). Our simulations show that in this case $\eta_0^2$ 
behaves as $2/3\la P_0\ra $ for the entire range of $\la P_0\ra$ as in the  case without autoregulation 
(see figure \ref{fig:etaauto}, black line with filled triangles).

Finally we consider the case where the synthesis of the protein is a stochastic process, but partitioning during cell  division is deterministic (see figure \ref{fig:autostoc}(b) for the protein number 
and figure \ref{fig:autostoc}(e) for corresponding concentration). 
In this case for small $\la P_0\ra$, autoregulation does not kick in and the system behaves effectively as an unregulated gene with  $\eta_0^2 = 1/3\la P_0\ra$. 
For large $\la P_0\ra$, both $\eta_0^2$ and the Fano factor $\eta_0^2\times \la P_0\ra$ are strongly reduced, as the synthesis becomes almost deterministic for 
large $\la P_0\ra$, where protein number fluctuations are suppressed by the negative autoregulation.
(see figure \ref{fig:etaauto} green line with filled circles). 

\begin{figure}
\hspace{-1.cm}
\includegraphics[width=8cm,height=6cm,angle=0.0]{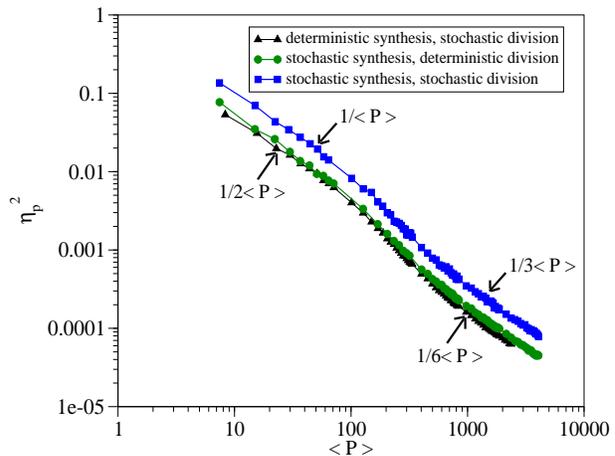}
\caption{ Noise strength of concentration $\eta_p^2$ as a function of the average protein number $\la P\ra$ 
(varied by varying the synthesis rate $\alpha_0$) for the different models with negative autoregulation. Averages are taken over whole division 
cycle. The parameters values are $V_0=2 \mu m^3$, $T= 50$ min, $K=100$ molecules$/\mu m^3$.}
\label{fig:conetaauto}
\end{figure}

In an experiment, one typically  looks at a population of cells, which are at different points in the 
division cycle. To address this situation we take averages of the protein number or concentration over many realization and over time through the full division  cycle instead of over different realizations, all taken directly after the division. In doing so, the
average protein concentration is a more relevant parameter than the average protein number, because the protein number increases two-fold over the division cycle. We therefore determine a noise parameter $\eta_p^2$ for the concentration. We plot simulation results for the
different cases in figure \ref{fig:conetaauto}. When both synthesis and division are stochastic (blue line with filled squares in figure \ref{fig:conetaauto}), $\eta_p^2$ retains the $1/\la P\ra$-behavior for small $\la P \ra$ (where the gene is effectively unregulated). For large $\la P \ra$,
that is in the presence of negative autoregulation, the noise is suppressed and $\eta_p^2$ shows a  $1/3\la P\ra$ behavior. The other two cases, namely stochastic synthesis and deterministic division and vice-versa exhibit the same dependence. For both the noise is increased compared to the earlier case where averages were taken 
immediately after division. Here $\eta_p^2$ goes as $\simeq 1/2\la P\ra$ for small $\la P\ra$ and $\simeq 1/6\la P\ra$ for large $\la P\ra$ 
(see figure \ref{fig:conetaauto}, black line with filled triangles and green line with filled circles). The latter results indicate that negative 
autoregulation suppresses noise from both sources (stochasticity of protein synthesis and stochastic partitioning during cell division). The observation 
that our results above (figure \ref{fig:etaauto}) only showed suppression of the synthesis noise and not of the partitioning noise, is due to taking 
averages directly after division, which leaves no time to compensate for variations of the protein concentration introduced during partitioning.

\section{Conclusions}
\label{sec:Conclusions}
In this paper, we have reviewed models  for constitutive (unregulated) protein synthesis that we have studied previously \cite{MaratheJSP11} and presented some new results for protein synthesis with negative autoregulation. In both cases, different model variants were considered to disentangle different sources of stochasticity, specifically stochastic synthesis of the protein and stochastic partitioning during cell division.  

The different models for unregulated gene expression show that there is no dominant source of stochasticity, as switching off one or the other source of noise leads to similar results (we note however that  'bursting' in protein synthesis, which we did not discuss here, can be dominant \cite{BergJTB78,OudenaardenPNAS01,MaratheJSP11}). We found similar behavior for the models with negative autoregulation and 
explicit linear volume growth, but for these models the relation between the noise parameter $\eta^2$ and the average protein number shows a crossover for protein concentrations at which autoregulation becomes important. Fluctuations are suppressed by negative autoregulation, as expected for such control systems and known from previous studies \cite{OudenaardenPNAS01,Becskei2000}. Specifically, our results show that fluctuations arising from both sources (stochastic synthesis and stochastic partitioning) are suppressed by negative autoregulation, as  shown by the approximately 3-fold reduction in the Fano factor for large values of $\la P \ra$, where the autoregulatory mechanism becomes active.



\begin{acknowledgments}
The authors would like to thank Veronika Bierbaum for useful discussions during the course of this study,
and Angelo Valleriani especially for calculation presented in the appendix.

\end{acknowledgments}

\appendix
\section{Calculation of moments for stochastic partitioning using generating functions}
\label{appA1}
\begin{figure}[tb]
\begin{center}
\includegraphics[width=.3\textwidth]{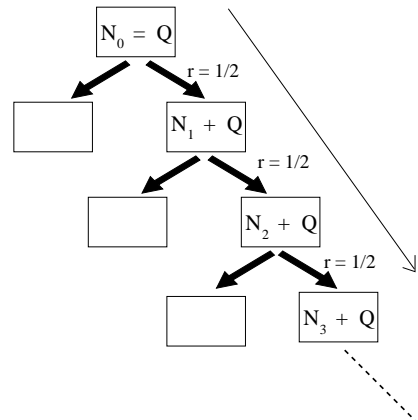}
\caption{Depiction of the model with deterministic synthesis and stochastic partitioning in cell division. We start with $N_0=Q$ particles,  
At every step (generation) the cell divides into two daughter cells and each proteins goes to one of the daughter cells with probability $r=1/2$. Between two divisions,  a constant number $Q$ of proteins is added, corresponding to a synthesis rate $\alpha=Q/T$. Only the rightmost branch of the tree diagram, i.e.\ one lineage of cells, is considered. }
\label{binsdetstoc}
\end{center}
\end{figure}

The case of deterministic addition of  $Q$ molecules during the cell division cycle and stochastic partitioning during cell division, depicted in figure \ref{binsdetstoc}, can be solved using the method of generating functions \cite{Fellerbook}. In this case the number of protein
molecules will follow a binomial distribution at the time of cell division with parameters $Q$ and $r$. Thus 
for the generating function we have
\bea
g(s)= (sr+1-r)^Q.
\eea
Let $X$ and $Y$ be two binomially distributed random variables. Let $g_{X|y}(v_1)$ be the generating function of the
variable $X$ conditioned to the fixed value of the variable $Y$ such that
\bea
g_{X|y}(v_1) = (v_1r+1-r)^y,
\eea
and let $g_Y(v_0)$ be the generating function for the variable $Y$, such that
\bea
g_{Y}(v_0) = (v_0r+1-r)^N.
\eea
Then we can write
\bea
g_{X}(v_1) &=& \sum_y g_{X|y}(v_1)Pr(Y=y)\nonumber\\ 
&=& (v_0r+1-r)^N\big\vert_{v_0=v_1r+1-r}.
\eea

Now let $N_1$ be the random number of molecules in box $1$ (see figurex \ref{binsdetstoc}). Its probability
distribution is given by:
\bea
\mathcal{P}(N_1=k|Q)=  {Q \choose k} r^{k}(1-r)^{Q-k}.
\eea
Now in the next generation we know that $Q$ particles are added and then distributed randomly between
daughter cells. Thus in the next generation we will need to divide $N_1+Q$ number of molecules with the distribution
\bea
\mathcal{P}(N_2=k|N_1+Q)=  {N_1+Q \choose k} r^{k}(1-r)^{N_1+Q-k}.
\eea
Then the unconditional probability $\mathcal{P}(N_2=k)$ is given by
\bea
\mathcal{P}(N_2=k)= \sum_{j=0}^{Q} \mathcal{P}(N_2=k|j+Q)~\mathcal{P}(N_1=j|Q).
\eea
Switching back to generating function and taking into account that $y=N_1+Q$, we find that
\bea
g_2(v_1)=\left( \prod_{j=0}^{1} (v_jr+1-r)\right)^Q\Bigg\vert_{v_0=v_1r+1-r},
\eea
which can be generalized to $m$ subdivisions leading to
\bea
g_m(v_{m-1})=\left( \prod_{j=0}^{m-1} (v_jr+1-r)\right)^Q\Bigg\vert_{v_j=v_{j+1}r+1-r}.
\eea 
This expression is enough to evaluate the moments. Let us call the product in the bracket $F_m$ then
$g_m= F_m^Q$ and let $v=v_{m-1} $. Also, we have 
\bea
\frac{dv_{m-j}}{dv}\ = r^{j-1}
\eea
for $1\leq j \leq m$ . This derivative is equivalent to
\bea
\frac{dv_{l}}{dv}\ = r^{m-1-l}
\eea 
for $0\leq l \leq m-1$. Then we have
\bea 
\frac{dg_{m}}{dv}\ = \sum_{j=0}^{m-1}\frac{Qr}{v_jr+1-r}\ \frac{dv_{j}}{dv}\  F_m^Q
\eea
and for the second derivative with $ \frac{dv_{j}^2}{dv^2}\ =0$, we have 
\bea
& &\frac{dg_{m}^2}{dv^2}\  =  F_m^Q \Bigg[-\sum_{j=0}^{m-1}\frac{Qr^2}{(v_jr+1-r)^2}\ \left(\frac{dv_{j}}{dv}\ \right)^2   \nn\\
 & & + \left(\sum_{j=0}^{m-1}\frac{Qr}{v_jr+1-r}\ \frac{dv_{j}}{dv}\ \right)
\left(\sum_{k=0}^{m-1}\frac{Qr}{v_kr+1-r}\ \frac{dv_{k}}{dv}\ \right)\Bigg],\nn\\ 
\eea
both of which need to be evaluated at $v_j=1$. We get
\bea
\frac{dg_{m}}{dv}\ \Big\vert_{v=1} = \frac{Qr}{1-r}\ (1-r^m),  
\eea
which in the limit $m\rightarrow\infty$ and $r=1/2$ leads to the average value $E[n|Q] = Q$. 
On the other hand the second derivative gives:
\bea
\frac{dg_{m}^2}{dv^2}\ \Big\vert_{v=1} = -\frac{Qr^2}{1-r^2}\ (1-r^{2m}) + \left(\frac{Qr}{1-r}\ (1-r^m)\right)^2,\nn\\
\eea
which for $m\rightarrow\infty$ and $r=1/2$ gives $E[n(n-1)|Q] = Q^2 -Q/3$. Finally we obtain the variance as:
\bea
Var[n|Q] = E[n(n-1)|Q]+E[n|Q]-E[n|Q]^2 = \frac{2}{3}\ Q,\nn\\
\eea
which gives $\eta^2 = 2/3Q$.

\end{document}